\documentclass[a4paper]{article}

\usepackage{INTERSPEECH2021}
\usepackage[utf8]{inputenc}
\usepackage{kotex}
\usepackage{kotex-logo}
\usepackage[noadjust]{cite}

\usepackage{amssymb}
\usepackage[hyphens]{url}
\usepackage[skip=1.5pt]{caption}
\usepackage{hyperref}

\title{FastPitchFormant: Source-filter based Decomposed Modeling \\ for Speech Synthesis}

\name{Taejun Bak, Jae-Sung Bae, Hanbin Bae, Young-Ik Kim, Hoon-Young Cho}
\address{Speech AI Lab, NCSOFT, Republic of Korea}
\email{\{happyjun,jaesungbae,bhb0722,youngik,hycho\}@ncsoft.com}
\begin{document}

\maketitle
\begin{abstract}
Methods for modeling and controlling prosody with acoustic features have been proposed for neural text-to-speech (TTS) models. Prosodic speech can be generated by conditioning acoustic features. However, synthesized speech with a large pitch-shift scale suffers from audio quality degradation, and speaker characteristics deformation. To address this problem, we propose a feed-forward Transformer based TTS model that is designed based on the source-filter theory. This model, called \emph{FastPitchFormant}, has a unique structure that handles text and acoustic features in parallel. With modeling each feature separately, the tendency that the model learns the relationship between two features can be mitigated. 
Owing to its structural characteristics, FastPitchFormant is robust and accurate for pitch control and generates prosodic speech preserving speaker characteristics. The experimental results show that proposed model outperforms the baseline FastPitch.
\end{abstract}
\section{Introduction}

Major objective of the neural text-to-speech (TTS) research is generating a natural voice that corresponds to a given sentence. Neural TTS models have become the mainstream in modern TTS research because they can synthesize natural sounding speech and comparable synthetic quality \cite{wang2017tacotron,shen2018natural,li2019neural}. In \cite{li2019neural}, a feed-forward Transformer (FFT) block was primarily  used 
to improve the synthetic quality of the mel-spectrogram.

Recently, non-autoregressive FFT-based TTS models have been proposed \cite{ren2019fastspeech,ren2020fastspeech,lancucki2020fastpitch}. Acoustic features, such as duration, pitch, and energy, were applied in the acoustic decoder of a TTS model and to predict the target mel-spectrogram. FastPitch \cite{lancucki2020fastpitch}, in particular, can control the prosody of synthesized speech at the fine-grained level by changing the character-level of the synthesized pitch values. 

In FastPitch \cite{lancucki2020fastpitch}, it was reported that FastPitch can generate the voice with manipulated pitch, which is referred to \emph{pitch-shift}, while preserving speaker characteristics. In our preliminary experiments, however, we observed that pitch expressiveness and speaker similarity decreased when the pitch values shifted with a deviation from the average pitch. We believe that this performance degradation is probably due to the structural vulnerabilities in FastPitch. The acoustic decoder in FastPitch handles not only text but pitch information together and generates speech from the pitch-conditioned text information. Therefore, the decoder is prone to learning the relationship between the text and pitch.

To separately handle text and prosodic information, an additional neural network, which was trained in an unsupervised manner, extracted the latent variables of the acoustic features 
\cite{prosody_tacotron,gst,zhang2019learning,hsu2018hierarchical,lee2019robust,sun2020generating,sun2020fully}. 
Then, the latent variables were applied in the acoustic decoder of the TTS model. In these studies, the prosody was controlled by changing the reference speech or modifying the extracted latent variables. However, because the latent variables were learned in the  unsupervised manner, the desired prosodic information may not be included in the latent variables.

Another approach uses the source-filter theory \cite{fant1970acoustic} which describes human speech production. Speech sounds are described as the responses of a sound source and a vocal tract filter in a source-filter model. The sound source and formant frequencies formulated by the vocal tract filter affect the fundamental frequency and phonation, respectively \cite{goldstein1973optimum,peterson1952control}. Several researchers have proposed singing voice synthesis models based on this approach  \cite{lee2019adversarially,lee2020disentangling}. However speech is the domain where the duration per character is short compared to singing, and the pitch also changes more frequently within a shorter time. In speech synthesis, the approach based on the source-filter theory have been applied in several researches for modeling waveform \cite{yoshimura1999simultaneous, Ai2020, wang2019neural}. However, to the best of our knowledge, the source-filter theory has not yet been applied in  neural TTS for generating the mel-spectrogram.

In this paper, we propose a non-autoregressive FFT-based TTS model based on the source-filter theory called FastPitchFormant. The main approaches in FastPitchFormant are (1) decomposed structure and (2) learning objective. With these approaches, FastPitchFormant can generate the mel-spectrogram using formant- and excitation-related representations which are separately modeled.
We evaluated the pitch controllability with several objective measurements. Furthermore, speech quality and speaker preservation of speech with pitch-shift were also evaluated using subjective listening tests. 
\section{FastPitchFormant}

Figure 1 depicts the FastPitchFormant structure. %
FastPitchFormant has four module types: (1) text encoder, (2) temporal predictor, (3) formant and excitation generators, and (4) spectrogram decoder. All components except temporal predictors consist of stacks of feed forward Transformer (FFT) blocks. The number of FFT blocks are six, four, and two, respectively.
The temporal predictors consist of 2 one-dimensional convolutional layers and predict ground-truth duration and pitch. For multi-speaker TTS, we applied a speaker embedding lookup table to obtain speaker embedding. The remainder of this section provides further details on each module type. 

\subsection{The Text Encoder and The Temporal Predictor}

The phoneme embedding vectors are represented by a phoneme sequence and a look-up embedding table with positional embedding. The phoneme embedding vectors pass through the text encoder, which then predict the hidden embedding. The hidden embedding is the input of two temporal predictors, the duration and pitch. The pitch embedding is obtained from the predicted pitch values passed through the one-dimensional convolutional layer. The hidden and pitch embedding are combined with the speaker embedding, respectively. The two representations are then discretely up-sampled and aligned with the predicted duration. We represent the up-sampled phoneme representation as $h \in\mathbb{R}^{D\times T}$, and the up-sampled pitch representation as $p \in\mathbb{R}^{D\times T}$, where $D$ is the dimension of the vectors, and $T$ is the total number of frames. $h$ and $p$ pass through the formant and excitation generators, respectively.

\subsection{The Formant and Excitation Generator}

We introduce formant and excitation generators into the model that are inspired by the source-filter theory \cite{fant1970acoustic}. The formant generator predicts the formant representation including formant-related information such as the linguistic information only using $h$. The excitation generator predicts the excitation representation including the excitation-related information such as the prosody using both $h$ and $p$. In our preliminary experiments, we observed that the pitch control accuracy is compromised when the excitation representation only utilizes $p$. To improve the pitch control accuracy, we applied a similar extension as that in \cite{selfattn_w_pos} to the self-attention mechanism. In first self-attention layer in the excitation generator, the attention matrix and query $Q$ are calculated as follows: 
\begin{equation}
    \text{Attention}(Q,K,V)=\text{softmax}(\frac{QK^T}{\sqrt{d}})V,
\end{equation}
\begin{equation}
    Q=W_Q(h+p)+b_Q,
    \label{q_att}
\end{equation}
\vspace{0pt}

\noindent where $K$ and $V$ are the matrices for the key and value in the self-attention mechanism, respectively, and $W_Q$ and $b_Q$ are the weight matrix and bias for the query, respectively. The effectiveness of this query extension is detailed in Section 3.3.1.

\begin{figure}[t]
  \centering
  \centerline{\includegraphics[width=1\linewidth]{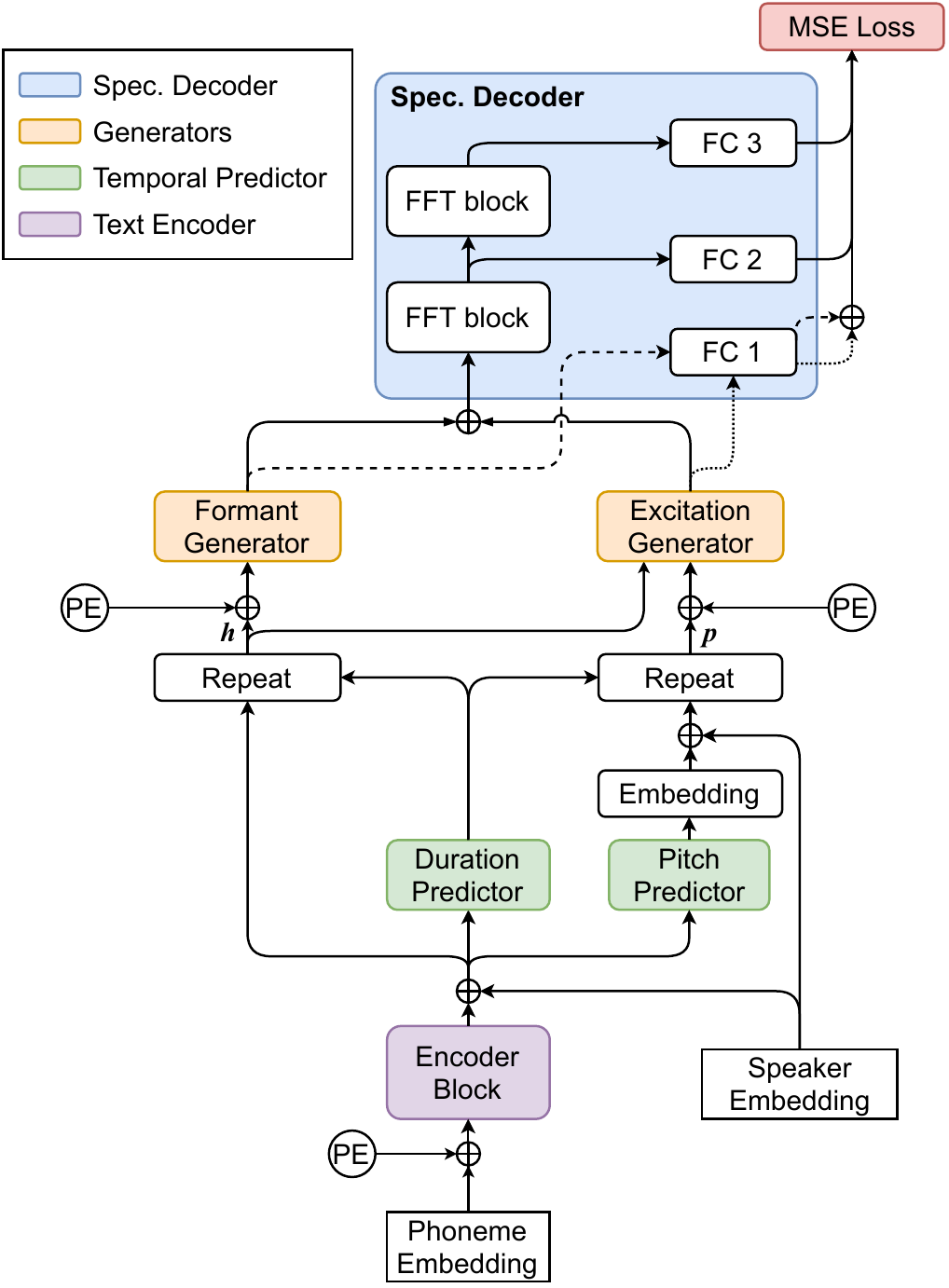}}
  \caption{Diagram of FastPitchFormant. Dashed and dotted lines represent formant and excitation representations, respectively.}
  \label{fig:fpf}
\end{figure}

\subsection{The Spectrogram Decoder}

The spectrogram decoder is comprised of two stacked FFT blocks and three fully connected (FC) layers. Each FC layer generates the target mel-spectrograms. 
The first spectrogram is generated by the summation of the projected formant and excitation representations through the first FC layer. The first FC layer is shared by two representations.
To produce the second and third mel-spectrograms, the summation of the formant and excitation representations is passed to the two stacked FFT blocks and then projected to mel-spectrogram by the second and third FC layers. In the source-filter theory, the source spectrum is multiplied by the vocal-tract filter. However, because our model handles log-scaled mel-spectrograms, we substitute the multiplication operation to summation. 
The outputs of each FC layer are used for the learning objective which includes all $L_2$ losses as an iterative loss in \cite{elias2020parallel}. Because of the iterative loss, the spectrogram decoder is trained to generate the final mel-spectrogram from the summation of the formant and excitation representations, while the two generators are trained to form those representations. In the inference stage, the mel-spectrogram from the third FC layer is the final output of FastPitchFormant.

\subsection{Learning objective}
The learning objective of FastPitchFormant is as follows:
\begin{equation}
    \mathcal{L}_{final}={\frac{1}{TM}}\sum^{3}_{i=1}\mathcal{L}_{spec_i}+\alpha\mathcal{L}_p+\beta\mathcal{L}_d,
\label{loss}
\end{equation}
where $M$ is the number of mel-spectrogram bins, $\mathcal{L}_{spec_i}$ is the $L_2$ loss between the target and $i$-th predicted mel-spectrogram from the $i$-th FC layer. $\mathcal{L}_{p}$ and $\mathcal{L}_{d}$ are $L_2$ loss between target and predicted pitch value and that of duration value, respectively. Note that any additional targets are not required to supervise the model to separately generate formant and excitation representations.

\section{Experiments}

\subsection{Dataset}

We used an internal Korean speaker dataset sampled at $22.05$ kHz, that contains $22$ h of speech from a female speaker and $17$ h of speech from a male speaker. One percent of the dataset was randomly selected for the test set.
We calculated an $80$-bin log-mel spectrogram with a fast Fourier transform size of $1024$, a hop size of $256$, and a window size of $1024$. We used a speech recognizer to extract a forced alignment with a phoneme sequence. Thus, we calculated phoneme-level pitch values by averaging the F0 values over every phoneme. The F0 values were extracted using the PRAAT toolkit \cite{praat}.

\subsection{Training Setup}

We trained FastPitch (FP) and FastPitchFormant (FPF) for up to $1000$k iterations using a mini-batch size of $16$ and the Adam optimizer \cite{kingma2014adam} with initial learning rate of $0.005$. The parameters of Adam optimizer were $({\beta}_1,{\beta}_2)=(0.5, 0.9)$, and ${\epsilon}=10^{-6}$. The learning rate decreased by half every $200$k iterations.
For the FFT block and temporal predictors of the models, we followed the same network architecture and hyperparameters as those in \cite{lancucki2020fastpitch}. VocGAN \cite{yang2020vocgan} was trained as the neural vocoder using a database containing approximately $40$ h of speech recorded by six speakers.

\subsection{Objective Evaluation}
To objectively evaluate the pitch controllability of the models, pitch-shifted speech was synthesized by FP and FPF. All audio were generated by manipulating the input pitch values of model in a semitone unit. The $\lambda$ semitone shifted pitch value, $f_\lambda$, can be calculated as follows:

\begin{equation}
    f_{\lambda}=2^{\frac{\lambda}{12}}\times f_{0},
\label{f0_shift}
\end{equation}
where $f_0$ is the original pitch value before shifted.
We then generated pitch-shifted speech of the test set with ground-truth duration and pitch for $\mathcal{\lambda} \in \{-8,-6,-4,0,4,6,8\}$. It implies that the original pitch shifts to $63\%$, $71\%$, $79\%$, $100\%$, $126\%$, $141\%$, and $159\%$ of its magnitude, respectively.

\setlength{\tabcolsep}{4pt}
\begin{table}[t]
\centering
    \caption{FFE (\%) results of pitch-shifted speech. The numbers in parentheses are the ratio varied from the original pitch.}
    \label{tab:FFE}
    \centering
    \begin{tabular}{l|cccccc}
        \toprule
                               & \multicolumn{6}{c}{\textbf{Pitch shift scale ($\lambda$)}} \\ \hline
        \textbf{Method}        & \begin{tabular}[c]{@{}c@{}}\textbf{-8}\\\scriptsize{(63\%)}\end{tabular} &           \begin{tabular}[c]{@{}c@{}}\textbf{-6}\\ \scriptsize{(71\%)}\end{tabular} &
        \begin{tabular}[c]{@{}c@{}}\textbf{-4}\\ \scriptsize{(79\%)}\end{tabular} &
        \begin{tabular}[c]{@{}c@{}}\textbf{+4}\\ \scriptsize{(126\%)}\end{tabular} &
        \begin{tabular}[c]{@{}c@{}}\textbf{+6}\\ \scriptsize{(141\%)}\end{tabular} & \begin{tabular}[c]{@{}c@{}}\textbf{+8}\\ \scriptsize{(159\%)}\end{tabular} \\ \hline
        \textit{FP (baseline)} & 44.90  & 32.81  & 21.36 & 15.59 & 25.60 & 37.10 \\ \hline
        \textit{FPF}           & 44.83  & 32.76  & 19.61 & 13.04 & 20.81 & 29.66 \\
        \textit{FPF w/o Q}     & 56.06  & 42.72  & 26.04 & 16.86 & 26.27 & 39.80 \\
        \bottomrule
    \end{tabular}
\end{table}

\subsubsection{Pitch Control Accuracy}
To evaluate the pitch control accuracy, we calculated the f0 frame error (FFE) \cite{chu2009reducing} between the extracted pitch values from the pitch-shifted speech generated using $f_{\lambda}$ and the shifted input pitch $f_{\lambda}$.
We also measured the FFE of FPF without the extension for query (FPF w/o Q), which is represented in Equation (\ref{q_att}), to evaluate its effectiveness. The results are listed in Table \ref{tab:FFE}. A low FFE value represents that the model can generate speech with the desired pitch. The results exhibited that FPF improved speech reproducibility with a wider range of pitch control compared to that of the baseline. When $\lambda$ was bigger than $0$, the difference between the FFE from FPF and FP was significant. 
In FPF w/o Q, the FFE was higher than that of the other two models. We observed that the formant generator took over most of the mel-spectrogram generation task as the number of training epochs increased in training FPF w/o Q. 

\begin{figure}[t]
  \centering
  \centerline{\includegraphics[width=0.8\linewidth]{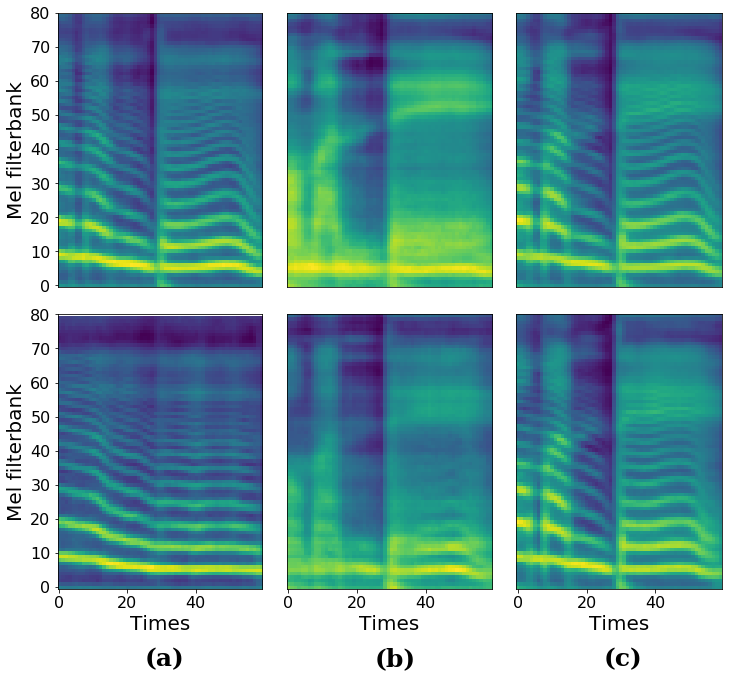}}
  \caption{Generated mel-spectrograms of (a) excitation representation, (b) formant representation and (c) final system output.
  Mel-spectorgrams of first and second rows are from FPF and FPF w/o Q, respectively.
  Inaccurate and undesired pitch contours are observed in the excitation and formant representations from FPF w/o Q.}%
  \label{fig:spectrogram}
\end{figure}

The mel-spectrogram examples from the excitation and formant representations are depicted in Figures \ref{fig:spectrogram}a and \ref{fig:spectrogram}b, respectively. They were generated by passing the excitation and formant representations through the spectrogram decoder individually.
The first row in Figure \ref{fig:spectrogram} shows that the formant and excitation generators in FPF were trained to model the action of the vocal cord and the vocal tract for generating speech. We conjecture that the separation comes from the difference in features exposed to each generator. In the early stage of training, prosodic-related parts such as pitch contours were generated first by the excitation generator handling the distribution of pitch that is a relatively low-level feature. As training proceeded, we can observe that the phonation had been formed gradually by the formant generator modeling linguistic features that are high-level feature.
In the formant representation from FPF w/o Q, we observed contours that were similar to the contours from the final mel-spectrogram and a pitch contour in the excitation representation was flat compared that of the FPF. Therefore, because of the extension, tasks for generating speech were properly distributed to generators. 

\begin{figure}[t]
  \centering
  \centerline{\includegraphics[width=0.9\linewidth]{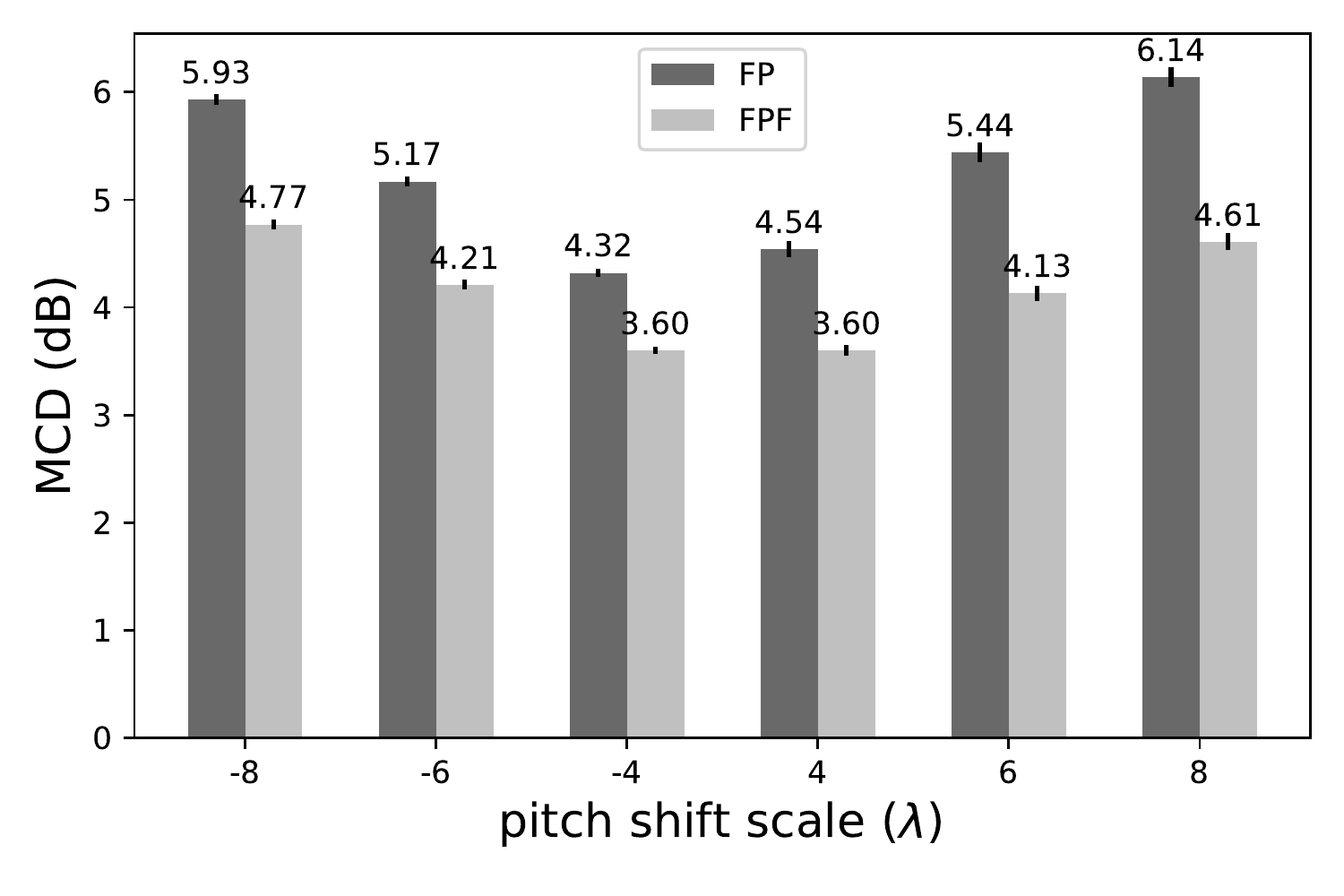}}
  \caption{MCD comparison between FP and FPF. with 95 \% Confidence Interval (CI). For the variation ratio from the average pitch in Hz unit, see Table \ref{tab:FFE}.}
  \label{fig:MCD}
\end{figure}

\subsubsection{Robustness of Pitch Control}

When the formant and excitation generator decompose speech into the formant and excitation representations well,
the spectral envelope of the speech with pitch-shift should be same as that of the the speech without pitch-shift. 
Therefore, we calculated the mel-cepstral distortion (MCD) \cite{kubichek1993mel} between speech with pitch-shift and speech without pitch-shift. Figure \ref{fig:MCD} illustrates the results of the MCD according to $\lambda$ in both cases of FP and FPF. 
FPF had lower MCD compared to FP for all $\lambda$.
It can be elucidated that FPF can synthesize less distorted speech compared to FP even with a significant pitch variance.

\begin{figure}[t]
  \centering
  \centerline{\includegraphics[scale=1.5,width=1\linewidth]{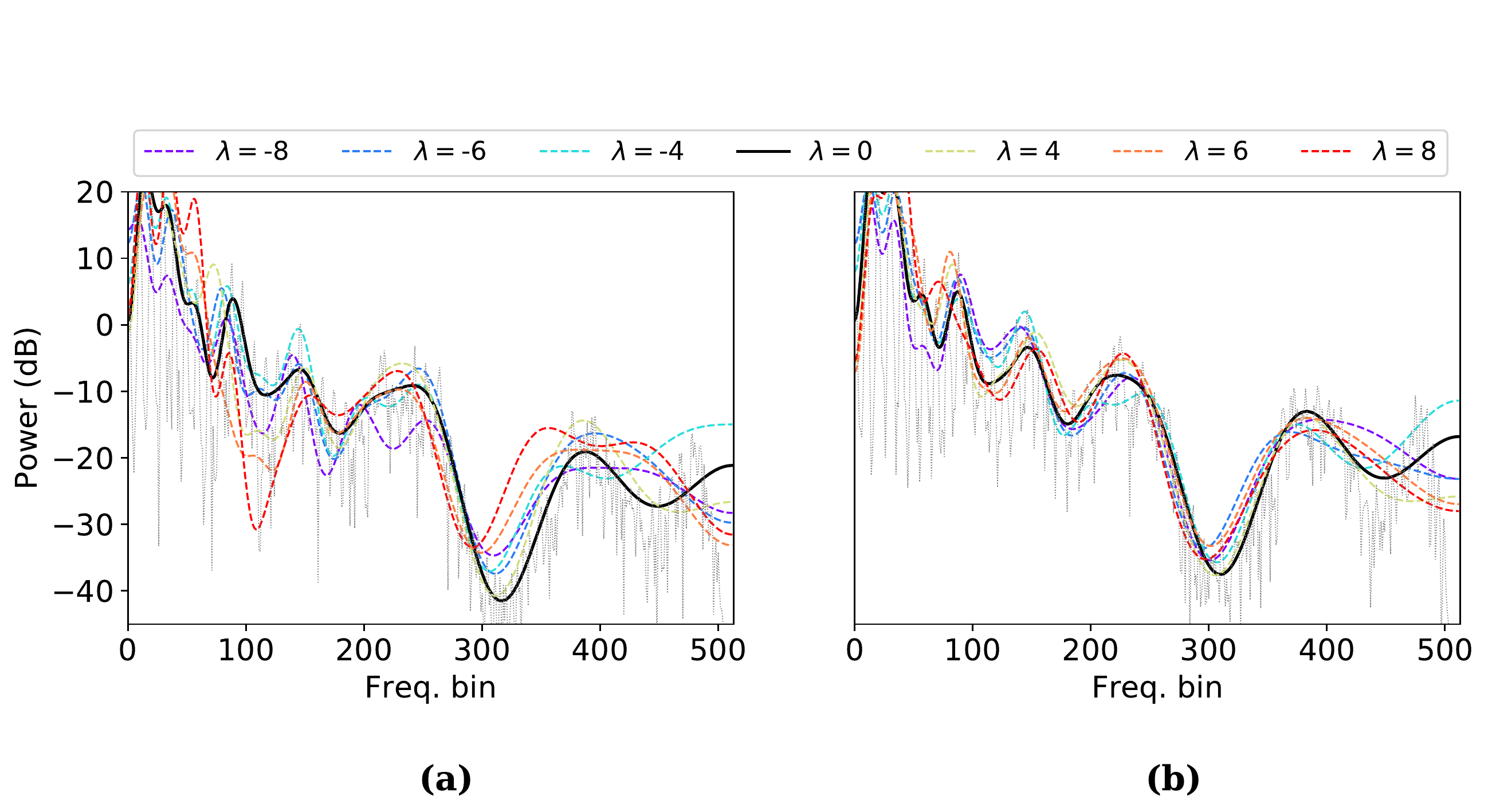}}
  \caption{Examples of spectral envelopes according to shifting pitch from (a) FP and (b) FPF for same speech frame. Black solid line is the spectral envelope of speech with $\lambda=0$, dashed lines are spectral envelopes for for different values of $\lambda$ with different colors and grey dotted lines are the power spectrum of speech. (Best viewed in color).}
  \label{fig:sp}
\end{figure}

We examined the changes in spectral envelopes for every $\lambda$ to visually confirm this hypothesis. Figure \ref{fig:sp} depicts the example of spectral envelopes in the same frame of synthesized speech from FP and FPF. For FPF, the spectral envelopes for every $\lambda$ appeared to maintain their original shape, while those from FP were distorted according to $\lambda$. This indicate that FPF can synthesize speech with pitch-shift which has more consistent pronunciation compared to FP because of its decomposed structure.

\begin{table}[t]
    \caption{MOS results of speech without pitch-shift with 95\% CI.}
    \centering
    \label{tab:MOS_norm}
    \begin{tabular}{lc}
        \toprule
        \textbf{Method}  & \multicolumn{1}{c}{\textbf{MOS}} \\ \midrule
        \textit{GT}      & \multicolumn{1}{c}{4.66 $\pm$ 0.09} \\
        \textit{GT (Mel+VOC)}  & \multicolumn{1}{c}{4.68 $\pm$ 0.09} \\ \midrule
        \textit{FP (baseline)}  & \multicolumn{1}{c}{4.08 $\pm$ 0.14}                     \\
        \textit{FPF} & \multicolumn{1}{c}{4.12 $\pm$ 0.14}                   \\
        \bottomrule
    \end{tabular}
\end{table}

\subsection{Subjective Evaluation}

For the subjective evaluation, we compared a mean opinion scores (MOS) and speaker preservation of pitch-shifted speech generated by FP and FPF. Pitch values were manipulated same as Equation (\ref{f0_shift}) with $\mathcal{\lambda} \in \{-8,-6,-4,4,6,8\}$. In addition, to evaluate MOS of speech without pitch-shift, ground-truth audio (GT) and audio generated by converting ground-truth mel-spectrogram to waveform with VocGAN (GT (Mel+VOC)) were compared together. All methods generated speech from the same input transcripts and predicted duration and pitch.

\subsubsection{Audio Quality}

Twenty samples from each model were randomly listed and evaluated, a total of $80$ samples. A total of $18$ native Korean speakers participated and were asked to score from $1$ to $5$ for each sample\footnote{Samples are available at \url{https://nc-ai.github.io/speech/publications/fastpitchformant}.}.

Table \ref{tab:MOS_norm} presents the MOS results of the pitch-shifted case and Table \ref{tab:MOS_shift} presents the MOS results of the not pitch-shifted case. We found that FPF results in MOS were comparable to those of the baseline in the not pitch-shifted case. In the pitch-shift case, FPF generated speech with close audio quality to that of the synthesized speech without pitch-shift, even when $|\lambda|=4$. As the magnitude of pitch shift scale was increased, the difference between the FP and FPF grew larger. We can therefore conclude that FPF generates speech exhibiting improved speech quality compared to that of FP even though the pitch was significantly shifted.

\setlength{\tabcolsep}{4pt}
\begin{table}[t]
    \caption{MOS results of pitch-shifted speech with 95\% CI. The numbers in parentheses are the ratio varied from the average pitch.}
    \centering
    \label{tab:MOS_shift}
    \begin{tabular}{l|cccccc}
        \toprule
                     & \multicolumn{6}{c}{\textbf{pitch shift scale ($\lambda$)}}                 \\ \hline
        \textbf{Method} &\begin{tabular}[c]{@{}c@{}}\textbf{-8}\\\scriptsize{(63\%)}\end{tabular} &       \begin{tabular}[c]{@{}c@{}}\textbf{-6}\\ \scriptsize{(71\%)}\end{tabular} &
        \begin{tabular}[c]{@{}c@{}}\textbf{-4}\\ \scriptsize{(79\%)}\end{tabular} &
        \begin{tabular}[c]{@{}c@{}}\textbf{+4}\\ \scriptsize{(126\%)}\end{tabular} &
        \begin{tabular}[c]{@{}c@{}}\textbf{+6}\\ \scriptsize{(141\%)}\end{tabular} & \begin{tabular}[c]{@{}c@{}}\textbf{+8}\\ \scriptsize{(159\%)}\end{tabular} \\ \hline
        \textit{\begin{tabular}[c]{@{}l@{}} FP (baseline)\\  $\quad\pm$C.I.\end{tabular}}  & \begin{tabular}[c]{@{}c@{}}1.69\\ 0.12\end{tabular} & \begin{tabular}[c]{@{}c@{}}2.57\\ 0.23\end{tabular} & \begin{tabular}[c]{@{}c@{}}3.38\\ 0.17\end{tabular} & \begin{tabular}[c]{@{}c@{}}3.67\\ 0.16\end{tabular} & \begin{tabular}[c]{@{}c@{}}2.92\\ 0.18\end{tabular} & \begin{tabular}[c]{@{}c@{}}1.97\\ 0.15\end{tabular} \\ \hline
        \textit{\begin{tabular}[c]{@{}l@{}} FPF\\ $\quad\pm$C.I.\end{tabular}} & \begin{tabular}[c]{@{}c@{}}2.77\\ 0.17\end{tabular} & \begin{tabular}[c]{@{}c@{}}3.38\\ 0.16\end{tabular} & \begin{tabular}[c]{@{}c@{}}3.55\\ 0.15\end{tabular} & \begin{tabular}[c]{@{}c@{}}3.8\\ 0.16\end{tabular} & \begin{tabular}[c]{@{}c@{}}3.33\\ 0.15\end{tabular} & \begin{tabular}[c]{@{}c@{}}2.74\\ 0.17\end{tabular} \\
        \bottomrule
    \end{tabular}
\end{table}

\begin{figure}[t]
  \centering
  \centerline{\includegraphics[width=0.95\linewidth]{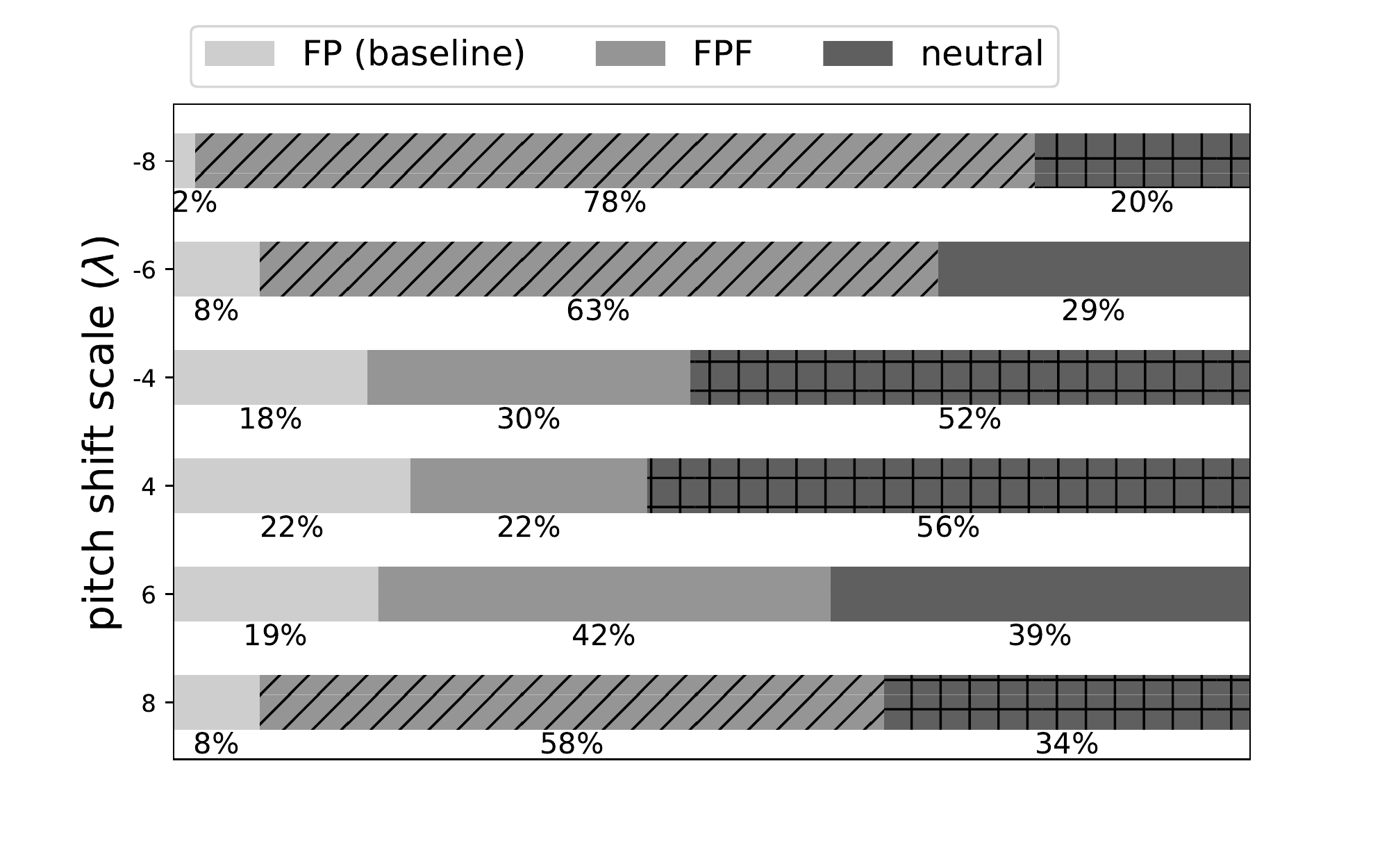}}
  \caption{Results of speaker similarity preference tests. For the variation ratio from the average pitch in Hz unit, see Table \ref{tab:MOS_shift}.}
  \label{fig:preference}
\end{figure}

\subsubsection{Speaker Preservation}
To evaluate the speaker preservation of speech with pitch-shift, we conducted a speaker similarity preference test. 
Participants were requested to select the speech more similar to original speaker voice from the FP and FPF samples. 
The same samples that were used in the MOS evaluation for pitch-shift case were used. The results are depicted on Figure \ref{fig:preference}. There was no significant difference in speaker similarity when the pitch-shift scale was relatively small ($|\lambda|\leq4$). However, most participants answered that samples from FPF preserved the speaker characteristics when $|\lambda|>4$ better than samples from FP. Thus, we verified that FPF can synthesize speech with pitch-shift, preserving speaker characteristics.

\section{Conclusion}
This study presents a non-autorgressive FFT-based TTS model called FastPitchFormant. Based on the source-filter theory, FastPitchFormant has the decomposed structure for separately handling text and acoustic features and generates speech from them. Objective results verified that FastPitchFormant has improved reproducibility of pitch and stability in pronunciation. Subjective results also showed that FastPitchFormant can synthesize speech which has better audio quality even for widely adjusted pitch values compared to FastPitch.

\bibliographystyle{IEEEtran}

\bibliography{mybib}

\begin{thebibliography}{10}
\providecommand{\url}[1]{#1}
\csname url@samestyle\endcsname
\providecommand{\newblock}{\relax}
\providecommand{\bibinfo}[2]{#2}
\providecommand{\BIBentrySTDinterwordspacing}{\spaceskip=0pt\relax}
\providecommand{\BIBentryALTinterwordstretchfactor}{4}
\providecommand{\BIBentryALTinterwordspacing}{\spaceskip=\fontdimen2\font plus
\BIBentryALTinterwordstretchfactor\fontdimen3\font minus
  \fontdimen4\font\relax}
\providecommand{\BIBforeignlanguage}[2]{{%
\expandafter\ifx\csname l@#1\endcsname\relax
\typeout{** WARNING: IEEEtran.bst: No hyphenation pattern has been}%
\typeout{** loaded for the language `#1'. Using the pattern for}%
\typeout{** the default language instead.}%
\else
\language=\csname l@#1\endcsname
\fi
#2}}
\providecommand{\BIBdecl}{\relax}
\BIBdecl

\bibitem{wang2017tacotron}
Y.~Wang, R.~J. Skerry-Ryan, D.~Stanton, Y.~Wu, R.~J. Weiss, N.~Jaitly, Z.~Yang,
  Y.~Xiao, Z.~Chen, S.~Bengio, Q.~Le, Y.~Agiomyrgiannakis, R.~Clark, and R.~A.
  Saurous, ``Tacotron: Towards end-to-end speech synthesis,'' vol. 2017, pp.
  4006--4010, 2017.

\bibitem{shen2018natural}
J.~Shen, R.~Pang, R.~J. Weiss, M.~Schuster, N.~Jaitly, Z.~Yang, Z.~Chen,
  Y.~Zhang, Y.~Wang, R.~Skerrv-Ryan, and R.~A. Saurous, ``Natural tts synthesis
  by conditioning wavenet on mel spectrogram predictions,'' in \emph{2018 IEEE
  International Conference on Acoustics, Speech and Signal Processing
  (ICASSP)}.\hskip 1em plus 0.5em minus 0.4em\relax IEEE, 2018, pp. 4779--4783.

\bibitem{li2019neural}
N.~Li, S.~Liuand, Y.~Liu, S.~Zhao, and M.~Liu, ``Neural speech synthesis with
  transformer network,'' in \emph{Proceedings of the AAAI Conference on
  Artificial Intelligence}, vol.~33, no.~01, 2019, pp. 6706--6713.

\bibitem{ren2019fastspeech}
Y.~Ren, Y.~Ruan, X.~Tan, T.~Qin, S.~Zhao, Z.~Zhao, and T.~Liu, ``Fastspeech:
  Fast, robust and controllable text to speech,'' in \emph{Advances in Neural
  Information Processing Systems}, vol.~32, 2019.

\bibitem{ren2020fastspeech}
Y.~Ren, C.~Hu, T.~Qin, S.~Zhao, Z.~Zhao, and T.~Liu, ``Fastspeech 2: Fast and
  high-quality end-to-end text-to-speech,'' \emph{arXiv preprint
  arXiv:2006.04558}, 2020.

\bibitem{lancucki2020fastpitch}
A.~{\L}a{\'n}cucki, ``Fastpitch: Parallel text-to-speech with pitch
  prediction,'' \emph{arXiv preprint arXiv:2006.06873}, 2020.

\bibitem{prosody_tacotron}
R.~J. Skerry-Ryan, E.~Battenberg, Y.~Xiao, Y.~Wang, D.~Stanton, J.~Shor,
  R.~Weiss, R.~Clark, and R.~A. Saurous, ``Towards end-to-end prosody transfer
  for expressive speech synthesis with tacotron,'' in \emph{Proceedings of the
  35th International Conference on Machine Learning}.\hskip 1em plus 0.5em
  minus 0.4em\relax PMLR, 2018, pp. 4693--4702.

\bibitem{gst}
Y.~Wang, D.~Stanton, Y.~Zhang, R.~S. Ryan, E.~Battenberg, J.~Shor, Y.~Xiao,
  Y.~Jia, F.~Ren, and R.~A. Saurous, ``Style tokens: Unsupervised style
  modeling, control and transfer in end-to-end speech synthesis,'' in
  \emph{Proceedings of the 35th International Conference on Machine
  Learning}.\hskip 1em plus 0.5em minus 0.4em\relax PMLR, 2018, pp. 5180--5189.

\bibitem{zhang2019learning}
Y.~J. Zhang, S.~Pan, L.~He, and Z.~Ling, ``Learning latent representations for
  style control and transfer in end-to-end speech synthesis,'' in \emph{ICASSP
  2019-2019 IEEE International Conference on Acoustics, Speech and Signal
  Processing (ICASSP)}.\hskip 1em plus 0.5em minus 0.4em\relax IEEE, 2019, pp.
  6945--6949.

\bibitem{hsu2018hierarchical}
W.~N. Hsu, , Y.~Zhang, R.~J. Weiss, H.~Zen, Y.~Wu, Y.~Wang, Y.~Cao, Y.~Jia,
  Z.~Chen, J.~Shen, and P.~Nguyen, ``Hierarchical generative modeling for
  controllable speech synthesis,'' in \emph{International Conference on
  Learning Representations}, 2019.

\bibitem{lee2019robust}
Y.~Lee and T.~Kim, ``Robust and fine-grained prosody control of end-to-end
  speech synthesis,'' in \emph{ICASSP 2019-2019 IEEE International Conference
  on Acoustics, Speech and Signal Processing (ICASSP)}.\hskip 1em plus 0.5em
  minus 0.4em\relax IEEE, 2019, pp. 5911--5915.

\bibitem{sun2020generating}
G.~Sun, Y.~Zhang, R.~J. Weiss, Y.~Cao, H.~Zen, A.~Rosenberg, B.~Ramabhadran,
  and Y.~Wu, ``Generating diverse and natural text-to-speech samples using a
  quantized fine-grained vae and autoregressive prosody prior,'' in
  \emph{ICASSP 2020-2020 IEEE International Conference on Acoustics, Speech and
  Signal Processing (ICASSP)}.\hskip 1em plus 0.5em minus 0.4em\relax IEEE,
  2020, pp. 6699--6703.

\bibitem{sun2020fully}
G.~Sun, Y.~Zhang, R.~J. Weiss, Y.~Cao, H.~Zen, and Y.~Wu, ``Fully-hierarchical
  fine-grained prosody modeling for interpretable speech synthesis,'' in
  \emph{ICASSP 2020-2020 IEEE International Conference on Acoustics, Speech and
  Signal Processing (ICASSP)}.\hskip 1em plus 0.5em minus 0.4em\relax IEEE,
  2020, pp. 6264--6268.

\bibitem{fant1970acoustic}
G.~Fant, \emph{Acoustic theory of speech production}.\hskip 1em plus 0.5em
  minus 0.4em\relax Walter de Gruyter, 1970, no.~2.

\bibitem{goldstein1973optimum}
J.~L. Goldstein, ``An optimum processor theory for the central formation of the
  pitch of complex tones,'' \emph{The Journal of the Acoustical Society of
  America}, vol.~54, no.~6, pp. 1496--1516, 1973.

\bibitem{peterson1952control}
G.~E. Peterson and H.~L. Barney, ``Control methods used in a study of the
  vowels,'' \emph{The Journal of the acoustical society of America}, vol.~24,
  no.~2, pp. 175--184, 1952.

\bibitem{lee2019adversarially}
J.~Lee, H.~S. Choi, C.~B. Jeon, J.~Koo, and K.~Lee, ``Adversarially trained
  end-to-end korean singing voice synthesis system,'' in \emph{Proc.
  Interspeech 2019}, 2019, pp. 2588--2592.

\bibitem{lee2020disentangling}
J.~Lee, H.~S. Choi, J.~Koo, and K.~Lee, ``Disentangling timbre and singing
  style with multi-singer singing synthesis system,'' in \emph{ICASSP 2020 -
  2020 IEEE International Conference on Acoustics, Speech and Signal Processing
  (ICASSP)}.\hskip 1em plus 0.5em minus 0.4em\relax IEEE, 2020, pp. 7224--7228.

\bibitem{yoshimura1999simultaneous}
T.~Yoshimura, K.~Tokuda, T.~Masuko, T.~Kobayashi, and T.~Kitamura,
  ``Simultaneous modeling of spectrum, pitch and duration in hmm-based speech
  synthesis,'' in \emph{Sixth European Conference on Speech Communication and
  Technology}, 1999.

\bibitem{Ai2020}
A.~Yang and L.~Zhen-Hua, ``Knowledge-and-data-driven amplitude spectrum
  prediction for hierarchical neural vocoders,'' in \emph{Proc. Interspeech
  2020}, 2020, pp. 190--194.

\bibitem{wang2019neural}
X.~Wang, S.~Takaki, and J.~Yamagishi, ``Neural source-filter-based waveform
  model for statistical parametric speech synthesis,'' in \emph{ICASSP
  2019-2019 IEEE International Conference on Acoustics, Speech and Signal
  Processing (ICASSP)}.\hskip 1em plus 0.5em minus 0.4em\relax IEEE, 2019, pp.
  5916--5920.

\bibitem{selfattn_w_pos}
P.~Shaw, Z.~Uszkoreit, and A.~Vaswani, ``Self-attention with relative position
  representation,'' in \emph{Proceedings of the 2018 Conference of the North
  {A}merican Chapter of the Association for Computational Linguistics: Human
  Language Technologies}, vol.~2, 2018, pp. 464--468.

\bibitem{elias2020parallel}
I.~Elias, H.~Zen, J.~Shen, Y.~Zhang, Y.~Jia, R.~Weiss, and Y.~Wu, ``Parallel
  tacotron: Non-autoregressive and controllable tts,'' \emph{arXiv preprint
  arXiv:2010.11439}, 2020.

\bibitem{praat}
P.~Boersma, ``Accurate short-term analysis of the fundamental frequency and the
  harmonics-to-noise ratio of a sampled sound,'' in \emph{Proc. institute of
  phonetic sciences}, 1993, pp. 97--110.

\bibitem{kingma2014adam}
D.~P. Kingma and J.~Ba, ``Adam: {A} method for stochastic optimization,'' in
  \emph{3rd International Conference on Learning Representations, {ICLR} 2015,
  San Diego, CA, USA, May 7-9, 2015, Conference Track Proceedings}, 2015.

\bibitem{yang2020vocgan}
J.~Yang, J.~Lee, Y.~Kim, H.~Cho, and I.~Kim, ``Vocgan: A high-fidelity
  real-time vocoder with a hierarchically-nested adversarial network,'' in
  \emph{Proc. Interspeech 2020}, 2020, pp. 200--204.

\bibitem{chu2009reducing}
W.~Chu and A.~Alwan, ``Reducing f0 frame error of f0 tracking algorithms under
  noisy conditions with an unvoiced/voiced classification frontend,'' in
  \emph{2009 IEEE International Conference on Acoustics, Speech and Signal
  Processing}.\hskip 1em plus 0.5em minus 0.4em\relax IEEE, 2009, pp.
  3969--3972.

\bibitem{kubichek1993mel}
R.~Kubichek, ``Mel-cepstral distance measure for objective speech quality
  assessment,'' in \emph{Proceedings of IEEE Pacific Rim Conference on
  Communications Computers and Signal Processing}, vol.~1.\hskip 1em plus 0.5em
  minus 0.4em\relax IEEE, 1993, pp. 125--128.

\end{thebibliography}

\end{document}